\documentclass[sigconf]{acmart} 

\AtBeginDocument{%
  \providecommand\BibTeX{{%
    \normalfont B\kern-0.5em{\scshape i\kern-0.25em b}\kern-0.8em\TeX}}}

\setcopyright{ceurws}
\copyrightyear{2021}
\acmYear{2021}
\acmDOI{}

\acmConference[ORSUM@ACM RecSys 2021]{4th Workshop on Online Recommender Systems and User Modeling, in conjunction with the 15th ACM Conference on Recommender Systems}{October 2nd, 2021}{Amsterdam, The Netherlands (Remote)}
\acmBooktitle{4th Workshop on Online Recommender Systems and User Modeling (ORSUM 2021), in conjunction with the 15th ACM Conference on Recommender Systems, October 2nd, 2021, Amsterdam, The Netherlands (Remote)}
\acmPrice{}
\acmISBN{}

\begin{document}

%\title{Active Learning for Recommender Systems: Learning in Batch or Sequentially?}
\title{Batch versus Sequential Active Learning for Recommender Systems}

\author{Toon De Pessemier}
\email{Toon.DePessemier@ugent.be}
\author{Sander Vanhove}
\email{Sander.Vanhove@ugent.be}
\author{Luc Martens}
\email{Luc1.Martens@ugent.be}
\affiliation{%
 \institution{Imec - WAVES - Ghent University}
 \streetaddress{iGent - Technologiepark 126}
 \city{9052 Gent}
 %\state{Arunachal Pradesh}
 \country{Belgium}}

\renewcommand{\shortauthors}{De Pessemier, et al.}

\begin{abstract}
Recommender systems have been investigated for many years, with the aim of generating the most accurate recommendations possible. However, available data about new users is often insufficient, leading to inaccurate recommendations; an issue that is known as the cold-start problem. A solution can be active learning. Active learning strategies proactively select items and ask users to rate these. This way, detailed user preferences can be acquired and as a result, more accurate recommendations can be offered to the user. % by actively selecting items and asking users to rate these, detailed user preferences can be acquired and as a result, more accurate recommendations can be offered to the users. 
In this study, we compare five active learning algorithms, combined with three different predictor algorithms, which are used to estimate to what extent the user would like the item that is asked to rate. In addition, two modes are tested for selecting the items: batch mode (all items at once), and sequential mode (the items one by one). Evaluation of the recommender in terms of rating prediction, decision support, and the ranking of items, showed that sequential mode produces the most accurate recommendations for dense data sets. Differences between the active learning algorithms are small. For most active learners, the best predictor turned out to be FunkSVD in combination with sequential mode.
\end{abstract}


%\keywords{datasets, neural networks, gaze detection, text tagging}


\maketitle

\section{Introduction}
Recommender systems are tools and techniques to assist users in the content selection process thereby coping with the problem of information overload. But the algorithms under the hood suffer from the cold start problem, i.e., the impossibility to provide accurate recommendations to new users or users with a small number of ratings. Moreover it has become essential to design recommender systems that are able to handle user feedback in real-time~\cite{al2018adaptive}. Active learning addresses these problems by proactively selecting items to be presented to the user in order to acquire her ratings and hence improve the output of the recommender. Active learner is a special case of machine learning that classifies data samples. It can acquire additional training samples by interactively querying a user for the correct labels of new data samples. The algorithm determines what data samples it is least certain about, and then asks the user to label these samples.

Active learning can also be used for recommender systems~\cite{rubens2015active}. The assumption is that some items can be rated by the user, because the user is familiar with the item and can assess whether it matches her preferences or the user has experienced the item in the past. In fact, these items might already be consumed by the user outside the content platform. For a movie recommender for example, the user might have seen a movie through another service, at a friend's house, in movie theater, etc., with the result that no rating has yet been provided in the video recommender. Although these items are not rated yet, the user is able to specify a rating, which is very useful data for the recommender. Therefore, an active learner will select some items whose ratings are very informative for the recommender. Next, the user is asked to rate these items, if possible, and the ratings are used for generating recommendations. But not all ratings have the same information value. The ratings of universally-liked items, i.e. items rated high by most people, are typically less informative. An additional rating for such a universally-liked item is probably again a high rating, but is not very informative for the user's preferences.

As a result, it is important to make a good selection of items that are offered to the user for rating. This is the task of the active learner. All those items can be selected at once, this is called the \emph{batch mode}. This selection of all these items is based on the ratings available before the active learning process starts. 
But those items can also be selected one by one, we refer to this as the \emph{sequential mode}. The selection of the first item is based on the ratings available before the active learning. The selection of the second item is based on these ratings, as well as the first rating obtained by the active learning. This continues in the same way for subsequent item selections. So, for the selection of the n-th item, the active learner has n-1 additional ratings to make a decision. 

The goal of this study is to investigate the difference between batch and sequential active learning and the impact on the accuracy of the final recommender, which is fed with rating data of the active learner. Five active learning algorithms are evaluated. To estimate how informative an item is, some of these learners need a recommender algorithm that predicts how the user will evaluate the item; we refer to this as a ``predictor algorithm''. The active learning algorithms are combined with three different predictor algorithms, and the impact of this predictor algorithm is evaluated. Active learning strategies are typically evaluated offline~\cite{carraro2020debiased}. In this study, two publicly available dataset are used for the evaluation.

%related work
Related work has shown that active learning approaches significantly increase the number of ratings acquired from the user and the recommendation accuracy~\cite{elahi2013personality}. Many different active learning strategies have been proposed in literature~\cite{elahi2014active}. Some of these early strategies are very simple, such as selecting the most popular items, the items with the highest variance, selecting random items, or a selection based on another aggregated statistic~\cite{rashid2002getting}. Later, some more advanced active learning strategies have been proposed~\cite{elahi2013personality}\cite{golbandi2011adaptive}. These advanced strategies analyze the available ratings of each individual user, and calculate which item rating would be the most informative to obtain as the next one. 

In addition, an active learning strategy has been proposed to combine the natural rating acquisition with an active learning algorithm~\cite{elahi2012adapting}. Elahi et al. compared the recommender accuracy obtained with various of these active learners, but they did not distinguish between selecting items in batch or sequentially~\cite{elahi2014active}. In online tests, the effect of the active learning technique on the user experience has been investigated~\cite{rashid2002getting}. However, according to our knowledge, the difference between active learning in batch and sequential mode has not yet been investigated for different algorithms.

The remainder of this paper is structured as follows. Section~\ref{AL} defines the active learning algorithms that were used in this study. The used recommendation algorithms are described in Section~\ref{recommender}. Section~\ref{experiment} discusses the experiment, with details about the data set, the data flow in the evaluation process, and the evaluation metrics. The results are reported in Section~\ref{results}. Finally, Section~\ref{conclusion} draws conclusions.

\section{Active Learner Algorithms}
\label{AL}

The goal of an active learner algorithm is to identify data set items, whose ratings are the most informative for a recommender system. Universally-liked items are typically less informative than controversial items in view of deriving user preferences and profiling. Two types of active learners can be distinguished: non-personalized and personalized active learners. 

Non-personalized active learners identify items that cover different user preferences, based on certain item metrics, such as rating variance, rating entropy, or item popularity, calculated with aggregated rating data of the community. Items with a high variance or entropy are informative because different users rate these items in a different way. Popular items are informative because they are useful for calculating user-user similarities, which are based on items rated by both users. Non-personalized active learners provide for each user the same set of ``most informative items'', regardless of the individual user's historical ratings. 

Personalized active learners, also called prediction-based active learners~\cite{elahi2014active}, identify items for which a rating would be informative, given the individual user's historical ratings. These personalized active learners provide for each user a personalized set of ``most informative items'', calculated based on the user's available ratings. For example, a rating for a specific item is less informative if a rating for a very similar item is already available. This personalized set of informative items is often calculated using an alternative recommendation algorithm, referred to as ``the predictor algorithm''. In this study, the focus is on personalized active learners, which are the most advanced ones.  

\subsection{Binary Prediction Active Learner} 
The Binary Prediction Active Learner is using an external predictor to rank items. 
The goal is to identify items that have a high probability that the user can rate them. Asking users to rate unknown items is not meaningful, since users will skip the ratings process or provide unreliable ratings. A recommendation algorithm can act as a predictor of user's item selections. Indeed, in an offline evaluation with datasets, a recommendation algorithm is often assessed on its ability to predict which items will be consumed by the user (e.g., with the precision and recall metric)~\cite{ricci2011introduction}. 

Binary Prediction first transforms the rating matrix to a binary matrix with the same number of rows and columns, by mapping null entries (missing ratings) to 0, and not-null entries (available ratings) to 1~\cite{elahi2013personality}. So, this binary matrix specifies which items have been rated by which users. Next, a recommendation algorithm is applied to this binary matrix, resulting in a prediction score for each item. This prediction score is used to rank the items. Items with a higher prediction score have a higher probability to be consumed by the user.

The idea of the Binary Prediction Active Learner is that the items with a higher score also have a higher probability that the user is familiar with these items, and as a result, can provide a rating. In this study, we utilized the Binary Prediction in combination with three different predictors described in Section~\ref{recommender}: FunkSVD, Item-Item Collaborative Filtering, and User-User Collaborative Filtering.

\subsection{Decision Tree Active Learner} 
\begin{figure}[htbp]
\centering%
\includegraphics[width=0.9\linewidth]{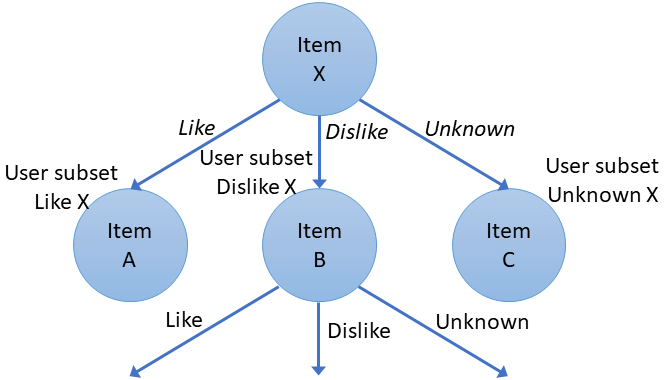}
\caption{Graphical representation of the Decision Tree Active Learner.}\label{DecisionTree}
   \Description[Decision Tree for Active Learning]{A Decision Tree in which nodes represent items. The branches partition the users according to their preference for the item: like, dislike, or unknown.}
\end{figure}

Decision trees can be used to decide which item of the data set to ask the user to rate. For this study, we adopted the efficient tree learning algorithm published by Yahoo~\cite{golbandi2011adaptive}. In this algorithm, each node of the decision tree evaluates user preference toward a certain item, and directs the user along a labeled edge to one of its subtrees based on the user's rating. There are three options corresponding to three edges: 
\begin{enumerate}
	\item The user likes the item by scoring it higher than the average rating value of the data set. 
	\item The user dislikes the item by scoring it lower than the average rating value of the data set. 
	\item The user does not know the item and is unable to rate it.
\end{enumerate}
Based on the user's ratings, the user follows a path from the root to a leaf that characterizes her. Thus after each split, the set of users is split into three partitions, one for each option. Subsequently, the algorithm continues with the partition of remaining users in the next node. Figure~\ref{DecisionTree} shows a graphical representation of this decision tree.

The item used in each node of the tree is selected based on an error metric, which is deﬁned as the squared deviation between the predicted ratings at the node and the true ratings of the users represented by the node. For more details regarding the construction of the tree, we refer to the original paper~\cite{golbandi2011adaptive}.

\subsection{Highest Prediction Active Learner} 

The Highest Prediction Active Learner applies a traditional predictor algorithm on the available data set and then asks the user to rate the items with the highest predicted score~\cite{elahi2014active}. The idea is that items with a high prediction score have a high probability to be familiar to the user, and can receive a rating. In case the item with the high prediction score receives a low rating from the user, this rating is a good correction of the user preferences. As traditional predictor, we used the same three algorithms as for the other active learners: FunkSVD, Item-Item and User-User Collaborative Filtering.

This active learner approach is well suited if the recommendation algorithm used for the final recommendations is very prone for the cold start problem and the predictor algorithm used in the active learner is not.

\subsection{Impact Analysis Active Learner}
Impact Analysis Active Learner searches for items that enable the recommender to make a prediction for other items that could not be predicted before~\cite{elahi2013personality}. A user-based collaborative filtering algorithm cannot generate a prediction for an item, if no similar user has rated it. Impact Analysis can identify user-item pairs for which the rating reveals additional user similarities. These user similarities can then be used to calculate rating predictions for other items.

\begin{figure}[htbp]
\centering%
\includegraphics[width=0.9\linewidth]{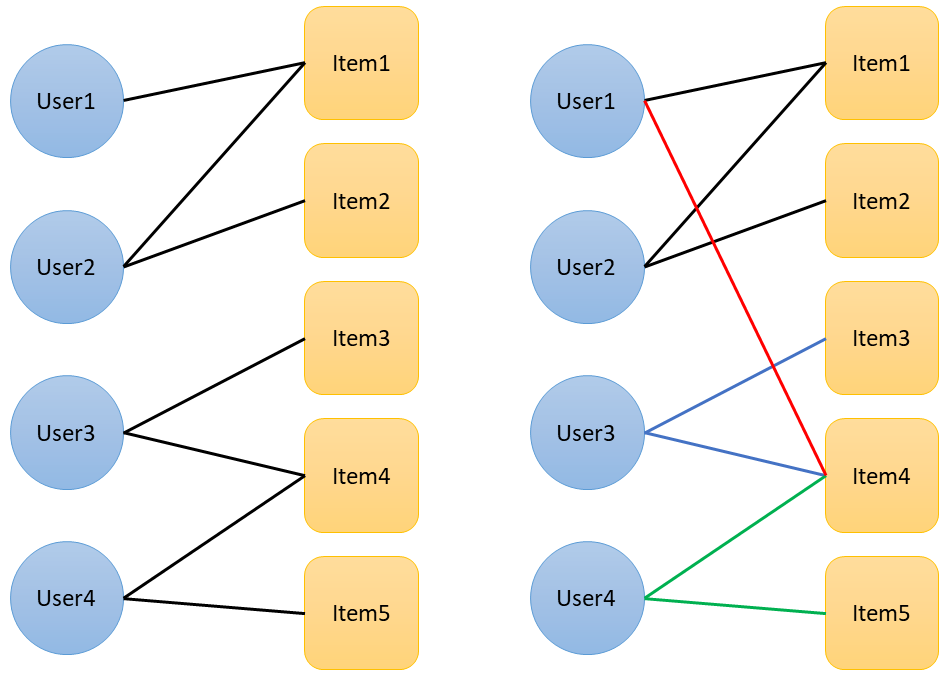}
\caption{Graphical representation of the ratings. According to Impact Analysis Active Learner, the rating (User1,Item4) is the most informative.}\label{ImpactAnalysis}
\Description[An example of Impact Analysis]{An example of a graph visualizing the ratings. It shows that the rating of User 1 for Item 4 creates two additional four-node paths.}
\end{figure}

Figure~\ref{ImpactAnalysis} shows the user-item pairs for which ratings are available on the left. On the right, the most informative rating according to Impact Analysis is indicated, it is the rating for the pair (User1,Item4). Users are represented as nodes (circles) and items are nodes as well (squares). If a user has rated an item, an edge between the two nodes is drawn. Collaborative filtering algorithms need four-node paths such as (User1-Item1-User2-Item2). Impact Analysis searches for edges that create the most additional four-node paths. In Figure~\ref{ImpactAnalysis}, a four-node path can be found from User1 to Item1 to User2 to eventually Item2. Because User1 and User2 both rated Item1, a collaborative filter is able to make a prediction about the preference of User1 for Item2, given that User2 is a neighbor of User1. 

Suppose that the active learner wants to gain additional information about User1, so that a collaborative filter can make a more informed decision. Which Item is the best to ask the user to rate? Item1 is already rated by User1. Item2 does not create any new four-node paths. The active learner could ask User1 to rate Item3 or Item5, because they both create one new four-node path, through User3 and User4 respectively. However, as indicated on the right side of Figure~\ref{ImpactAnalysis}, the most informative rating is for Item4. A rating of User1 for Item4 creates two new four-node paths for User1: (User1-Item4-User3-Item3) and (User1-Item4-User4-Item5). If User1 is able to rate Item4, the collaborative filter is able to make a prediction about Item3 and Item5.

\subsection{Lowest Prediction Active Learner}

The Lowest Prediction Active Learner is a variation on the Highest Prediction Active Learner~\cite{elahi2014active}. As indicated by the name, this learner asks the user to rate the items with the lowest prediction score. The probability that the user can rate the item might be lower, since the user dislikes the item according to the predictor. In the movie domain for example, it is likely that the user has not yet seen a movie if it does not fit into his sphere of interest. As a result, the user might be unable to rate this movie.

The advantage of this active learner is that the recommender gets some negative feedback. In general, users are more likely to rate the items they like, than items they dislike. Since users typically chose for items they think they will like, the majority of the ratings in data sets is often positive. This problem is also referred to as the self-selection bias. As a result, the recommender has difficulties in learning the difference between what a user dislikes, and what the user has not yet tried.

\section{Recommendation Algorithms}
\label{recommender}
The following three recommendations algorithms are used as predictor algorithm in the active learners.
\subsection{User-User Collaborative Filtering}
The User-User Collaborative Filtering algorithm calculates recommendations for the target user by identifying users that are similar in terms of rating behavior, also called neighbors~\cite{ricci2011introduction}. The target user's preference for unrated items is estimated based on the ratings of these similar users for these unrated items. A standard collaborative filtering algorithm, as described in~\cite{ricci2011introduction} has been implemented with the following parameters. Rating predictions are based on a minimum of 10 neighbors (to ensure accurate predictions) and maximum 30 neighbors (since more neighbors might not make the predictions more accurate, but increase the computational load and add more noise). For each of these neighbors, a minimum similarity value of $0.15$, calculated with the Pearson correlation, is required. 

\subsection{Item-Item Collaborative Filtering}
Item-Item Collaborative Filtering tries to identify similar items in terms of their ratings. Recommendations are calculated based on the similarities between items that received a high rating from the target user and items not yet rated by this target user. The standard algorithm is implemented with the following parameters. An item should have at least 1 neighbor, and a maximum of 20 neighboring items is used for each item. The similarity value of a neighboring item should be higher than $0.0$. This thresholds is rather low in order to find enough similar items for each item. In general, item-item similarities are more stable than user-user similarities, therefore less neighbors are used for item-item collaborative filtering.

\subsection{Funk SVD}
Funk SVD is a recommedation algorithm based on matrix factorization. Instead of an exact mathematical approach, it uses an iterative optimization based on gradient descent~\cite{funk2006netflix}. It is an effective algorithm that was successfully practiced by many others~\cite{koren2008factorization}. In our implementation, the amount of features per item is set to $25$, the number of iterations is set to $100$, and the learn rate is $0.001$.

\section{Experimental Setup}
\label{experiment}
The active learners are evaluated through an extensive experiment.
\subsection{Data set}
The evaluation is performed based on the Jester and the BookCrossing data set. The Jester data set is a rating set comprised of ratings about jokes, published by Goldberg et al. from the University of California~\cite{goldberg2001eigentaste}. Three different versions of the data set have been published. In this study, the data set is used that contains ratings for 100 jokes from 73,421 users summing to a total of 4.1 million ratings collected between 1999 and 2003. These rating are ranging from -10.0 to +10.0 using a continuous scale. Each user in this data set has rated at least 15 jokes and the majority rated more than 35 jokes. So, compared to other data sets, this data set is rather dense. This allows the recommender to accurately learn user preferences, which is the reason for choosing this dataset. 

The BookCrossing dataset~\cite{ziegler2005improving} contains 278,858 users (anonymized but with demographic information) providing 1,149,780 ratings (explicit / implicit) about 271,379 books. Ratings are expressed on a scale from 1 to 10. This data set is very sparse, which makes it more difficult to learn user preferences.

\subsection{Data Flow}
%active learner. 
The active learners of Section~\ref{AL} have been evaluated, and the ones that need a predictor are tested in combination with the algorithms of Section~\ref{recommender}. As final recommender algorithm, used to generate the recommendations for the users, the same algorithms of Section~\ref{recommender} were chosen.  %Results are shown for the User-User Collaborative filtering and FunkSVD algorithm. %This algorithm often suffers from the cold start problem, and is perfect to combine with an active learning algorithm. 
For each active learner, we evaluate the batch as well as the sequential mode. 
%
%data split & item selection
The evaluation is performed in the following subsequent phases:
\begin{enumerate}
	\item \emph{Select a user.} This user's ratings are removed from the training set and stored in a control data set. Except one rating of this user, randomly chosen, remains in the training set, and is used as input for the active learner. This simulates a new user, who has rated one item. %this rating is 4 or more on a five point rating scale
	\item \emph{Use the active learner.} %to get the most informative ratings from the user. 
	\begin{enumerate}
		\item \emph{Train the active learner.} The current training set is used.
		\item \emph{Select an item.} The active learner is used to calculate the next item that is the most informative for the user. 
		\item \emph{Add the rating to the training set.} If the rating for the selected user-item pair exists in the control set, it is marked as available and added to the training set. Otherwise, this rating is marked as missing.   
		\item \emph{Repeat item selection.} Items are selected until 6 ratings of this user could be added to the training set. (6 ratings + 1 initial rating showed to be a good number for generating recommendations for new users.) 
		\begin{itemize}
			\item In Batch mode $\Rightarrow$ go to 2b.
			\item In Sequential mode, and the rating is marked as available in 2c $\Rightarrow$ go to 2a.
			\item In Sequential mode, and the rating is marked as missing in 2c $\Rightarrow$ go to 2b.
		\end{itemize}
	\end{enumerate}
	\item \emph{Run the recommender.} Use the training set that includes the learned ratings from 2c to generate recommandations for the user. 
	\item \emph{Repeat user selection.} This is repeated until 100 users are evaluated $\Rightarrow$ go to 1.
\end{enumerate}

\subsection{Evaluation Metrics}
The accuracy of the final recommendations is evaluated using three commonly-used metrics, Root Mean Square Error (RMSE), Precision, and Normalized Discounted Cumulative Gain (nDCG). RMSE is a rating prediction metric that measures the accuracy of an algorithm based on the error between the predicted ($\hat{r}_{ui}$) and real ($r_{ui}$) rating value of a user $u$, for an item $i$. It squares the individual errors, calculates the mean value over all ratings ($n$ = number of ratings), and takes the square root of the mean to bring this to the scale of the ratings. The lower the value of the RMSE the better the rating prediction, where a value of zero stands for the perfect prediction. 

\begin{equation}
    RMSE = \sqrt{\frac{1}{n} \sum_{u \in U}\sum_{i \in I}(\hat{r}_{ui}-r_{ui})^2}
\end{equation}

Precision is a decision support metric that measures the percentage of recommended items that are relevant. The higher the precision, the better the recommendations. A perfect selection of recommendations will obtain a precision value of 1.

\begin{equation}
    Precision = \frac{\#items\;relevant\;and\;recommended}{\#items\;recommended} 
\end{equation}

The nDCG is a ranking metric that measures whether the algorithm ranks the items in the same order as the user. To calculate the nDCG, first the DCG of a list of recommendations is calculated. This is the discounted cumulative gain. It accumulates the gain of a list of items and multiplies the gain with a discount factor based on that item's position in the list.
\begin{equation}
    DCG(L) = \sum_{i \in L} gain(i) * disc(i)
\end{equation}

Where $L$ is the ordered list of items offered to the user as recommendations. As common practice for data sets with ratings, the gain is defined as the rating the user $u$ gave to item $i$:
\begin{equation}
    gain(i) = r_{ui}
\end{equation}
The discount factor for each item is proportional to its position within the list and defined as follows:
\begin{equation}
    disc(i) = \frac{1}{log_2(rank(i) + 1)}
\end{equation}
Where $rank(i)$ is the position of the item within the list. Because of this discount factor, items that are presented first are more important than items at the back of the list. 

This mimics the behavior of real users in the way that users' attention will be more drawn to and focused on the first items of the list. Gradually, their attention typically weakens over time towards the end of the list.
After calculating the DCG of the proposed recommendation list, it can be normalized as follows:
\begin{equation}
    nDCG(L)=\frac{DCG(L)}{DCG(L_{ideal})}
\end{equation}
Where $L_{ideal}$ is the ideal list to present to the user. This ideal list is a list of items, ordered according to the user's ratings from highest to lowest rating. These ratings can be available in the test set (or the control data set in our experiment) for evaluation. Of course, this ideal list is limited to the ratings that are available for the user and will almost never be the true ideal list for that user. The normalized discounted cumulative gain has a value between 0 and 1, the higher the better.

\section{Results}
\label{results}

\begin{figure*}[htbp]
\centering%
\includegraphics[width=0.98\linewidth]{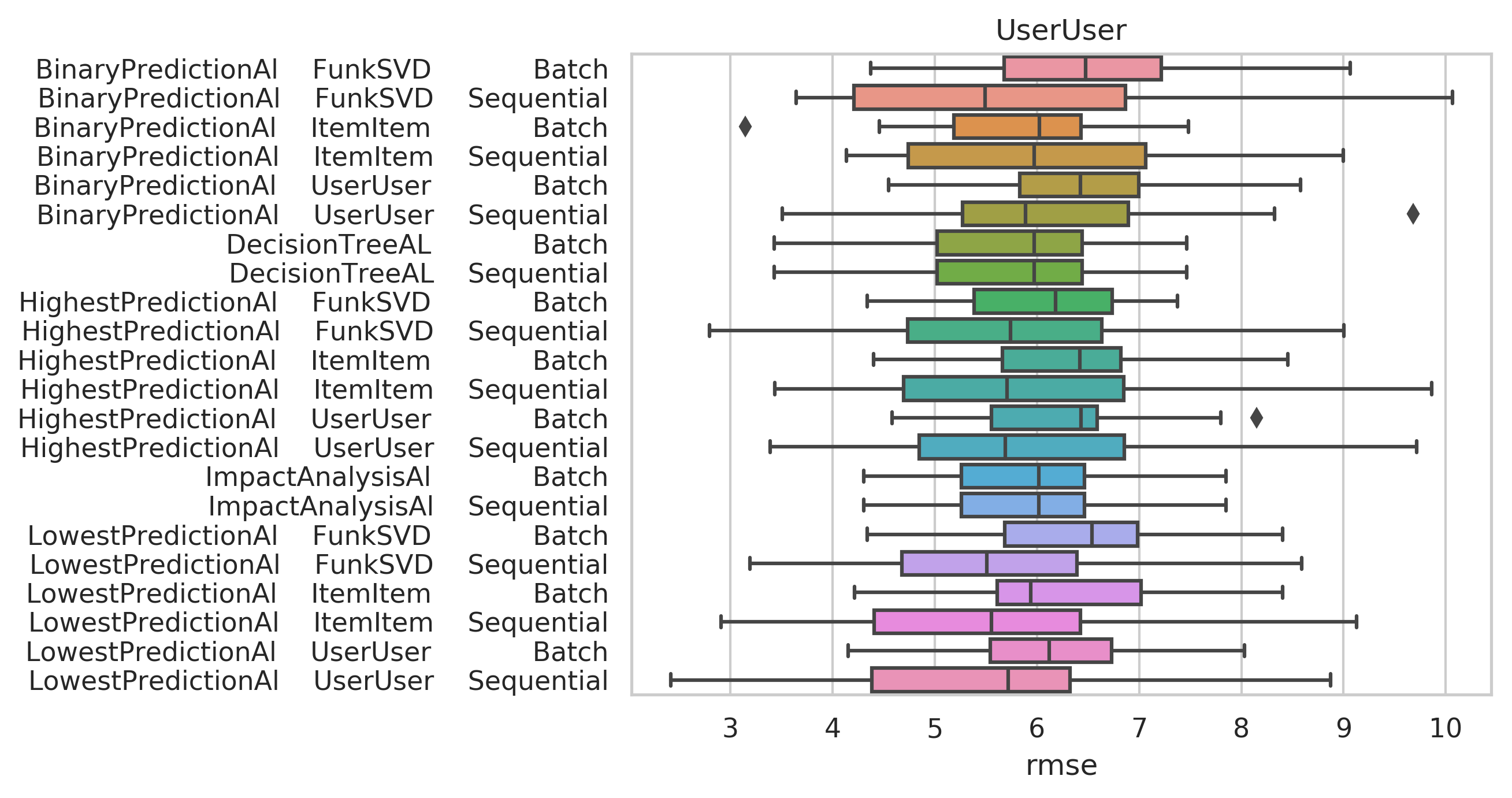}
\caption{RMSE value of user-user collaborative filtering combined with active learners with different algorithms on the Jester dataset. Lower is better.}\label{Jester-pal-UserUser-rmse}
\Description[Box-plots of the RMSE obtained for different active learners with the Jester dataset]{Box-plots of the RMSE obtained for different active learners  with the Jester dataset, showing that the median value in sequential mode is lower or the same than the median value in batch mode.}
\end{figure*}

Figure~\ref{Jester-pal-UserUser-rmse} shows the RMSE of the User-User Collaborative Filtering algorithm that is fed with rating data as determined by the active learner. The results are based on the Jester dataset. So for the main recommender, User-User Collaborative Filtering is always used in this graph. For the predictor algorithm that is part of the active learner, different options are tested (FunkSVD, UserUser, and ItemItem). The lower the RMSE, the better the algorithm can predict the ratings, and typically, the better the recommendations. The box plots show the minimum, first quantile (25\%), median, third quantile (75\%) and maximum value of the RMSE for all users in the dataset. 

Sequential active learning is clearly better than batch active learning. In sequential mode, informative items are selected one by one, and the ratings are moved from the control set to the training set. The selection of the most informative item is based on the available historical ratings (i.e. the original training set), as well as the ratings for the previously selected items that are informative for the recommender (i.e. ratings moved from control set to training set). So, sequential active learning can adjust the selection of informative items, one by one. In batch mode, all the informative items have to be selected at once, based only on the historical ratings. But ratings for these informative items cannot be taken into account. So, batch active learning has less information to make a decision, and as a result, it makes a suboptimal selection of informative items.

Comparing the different active learning algorithm shows that the Lowest Prediction Active Learner has the best results. But the difference with other active learners such as Highest Prediction and Binary Prediction are small. 

When we investigate the influence of the predictor used for the active learning algorithm (FunkSVD, ItemItem, or UserUser), we do not see major differences in the accuracy of the final recommendations. But FunkSVD combined with sequential mode gives consistently good results for all active learners. All these predictors are able to distinguish between the items users like and items they dislike. Minor differences in the predictions of these algorithms have only a limited impact on the final recommendations. 

\begin{figure*}[htbp]
\centering%
\includegraphics[width=0.98\linewidth]{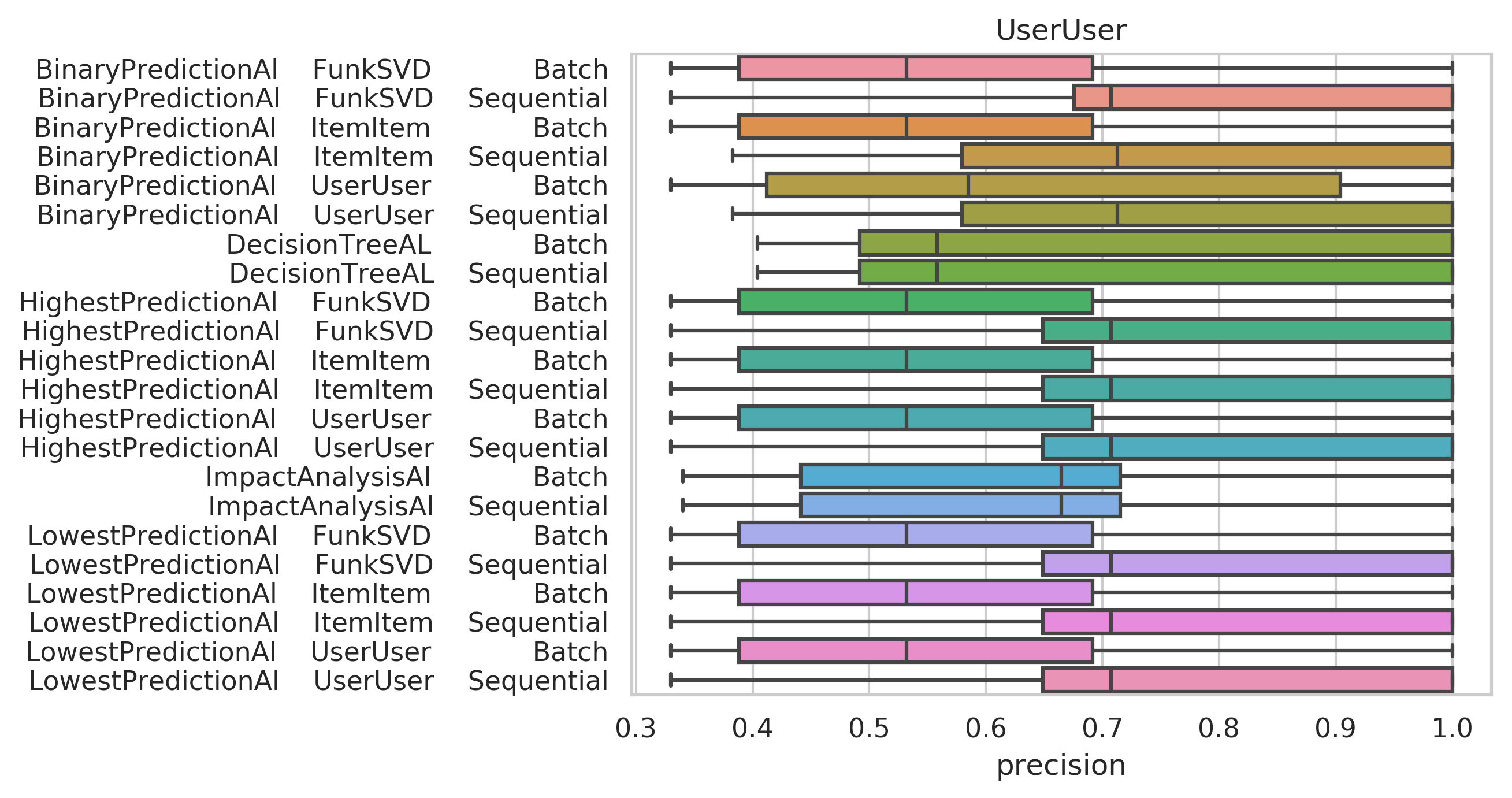}
\caption{The precision of user-user collaborative filtering combined with active learners with different algorithms on the Jester dataset. Higher is better.}\label{Jester-pal-UserUser-precision}
\Description[Box-plots of the Precision obtained for different active learners with the Jester dataset]{Box-plots of the precision obtained for different active learners with the Jester dataset, showing that the median value in sequential mode is higher or the same than the median value in batch mode.}
\end{figure*}

Figure~\ref{Jester-pal-UserUser-precision} shows the precision of the User-User Collaborative Filtering algorithm combined with different active learners for the Jester dataset. Again, a clear difference between sequential and batch mode can be recognized. Sequential mode obtains significantly better results in terms of precision. The differences between the active learning algorithms are small. There is no clear winner of the active learning algorithms.

\begin{figure*}[htbp]
\centering%
\includegraphics[width=0.98\linewidth]{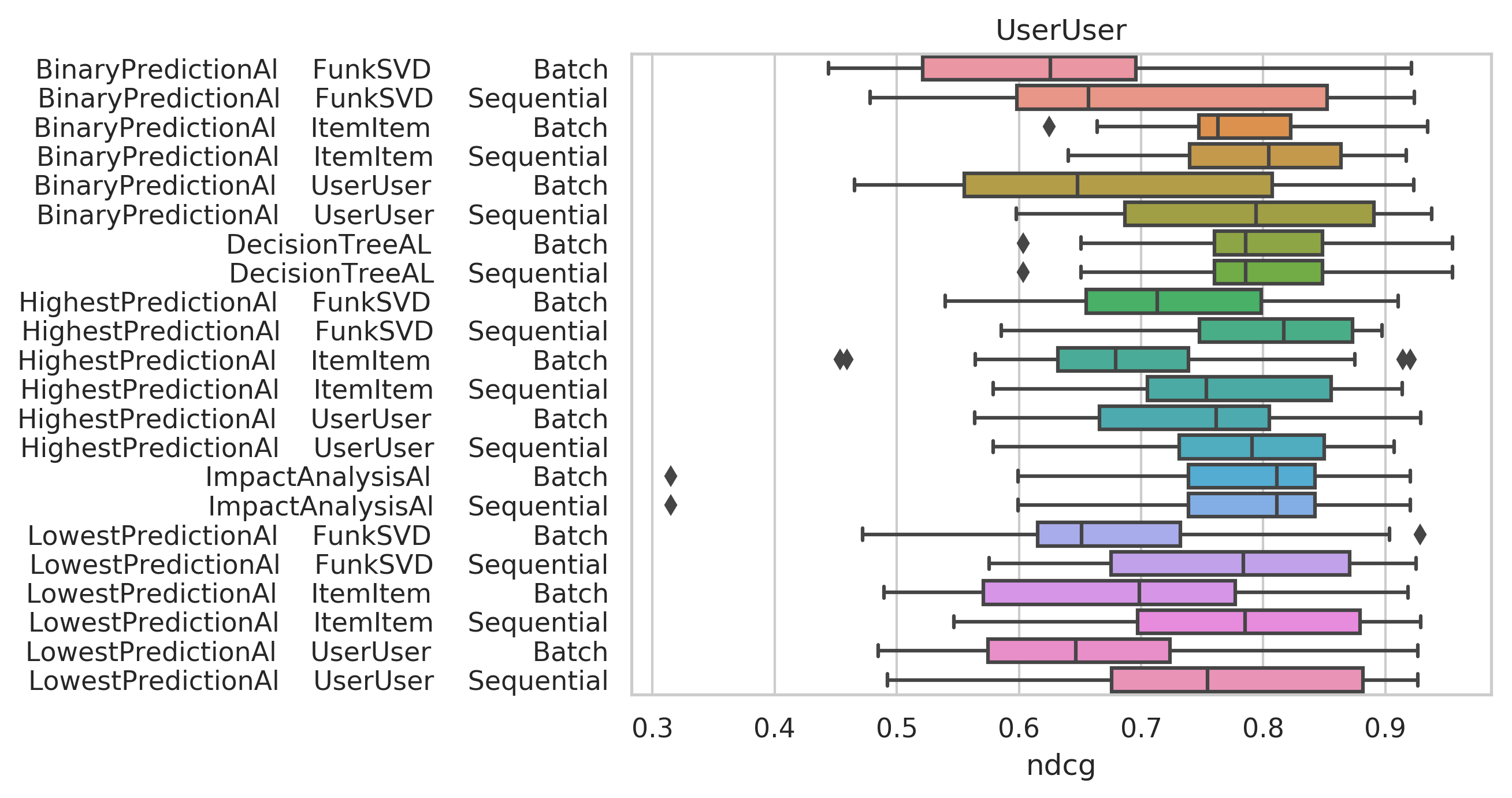}
\caption{nDCG value of user-user collaborative filtering combined with active learners with different algorithms on the Jester dataset. Higher is better.}\label{Jester-pal-UserUser-ndcg}
\Description[Box-plots of the nDCG obtained for different active learners with the Jester dataset]{Box-plots of the nDCG obtained for different active learners with the Jester dataset, showing that the median value in sequential mode is higher or the same than the median value in batch mode.}
\end{figure*}

Figure~\ref{Jester-pal-UserUser-ndcg} shows the nDCG of the User-User Collaborative Filtering algorithm combined with different active learners for the Jester data set. The higher the nDCG, the better the algorithm ranks the items, and typically, the better the recommendations. From these box plots, similar conclusions can be derived. More accurate recommendations are achieved using sequential active learning than using batch active learning. The differences in rank accuracy for the various active learners are small; but the best results (in terms of nDCG) are obtained with the Impact Analysis active learner. FunkSVD as predictor combined with sequential mode yields very accurate recommendations; except for the Binary Prediction active learner, this was not the best combination. 

\begin{figure*}[htbp]
\centering%
\includegraphics[width=0.98\linewidth]{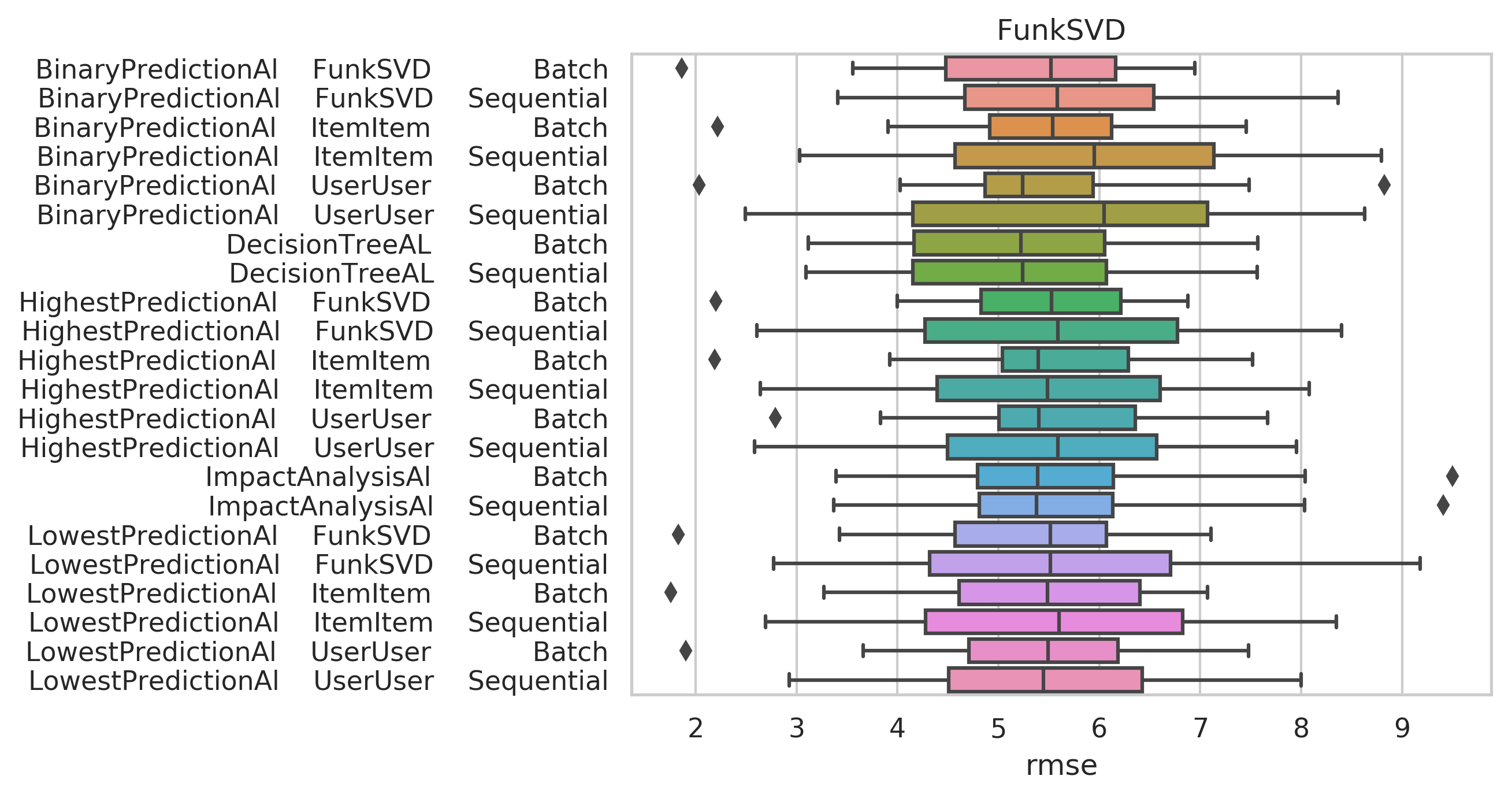}
\caption{RMSE value of the FunkSVD algorithm combined with active learners with different algorithms on the Jester dataset. Lower is better.}\label{Jester-pal-FunkSVD-rmse}
\Description[Box-plots of the RMSE obtained for different active learners with the Jester dataset]{Box-plots of the RMSE obtained for different active learners with the Jester dataset, showing that the median value in sequential mode is lower or the same than the median value in batch mode.}
\end{figure*}

\begin{figure*}[htbp]
\centering%
\includegraphics[width=0.98\linewidth]{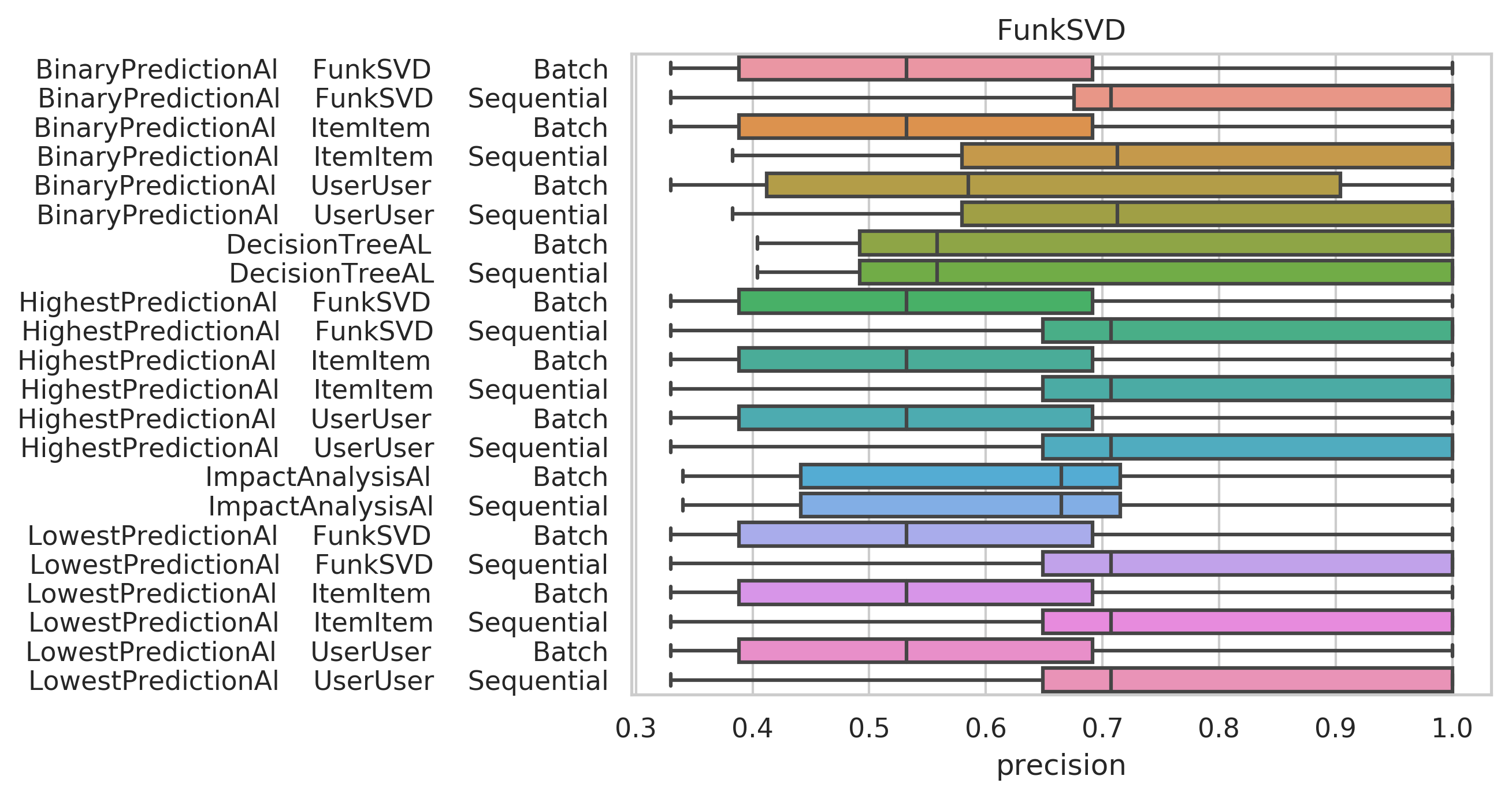}
\caption{The precision of the FunkSVD algorithm combined with active learners with different algorithms on the Jester dataset. Higher is better.}\label{Jester-pal-FunkSVD-precision}
\Description[Box-plots of the Precision obtained for different active learners with the Jester dataset]{Box-plots of the precision obtained for different active learners with the Jester dataset, showing that the median value in sequential mode is higher or the same than the median value in batch mode.}
\end{figure*}

\begin{figure*}[htbp]
\centering%
\includegraphics[width=0.98\linewidth]{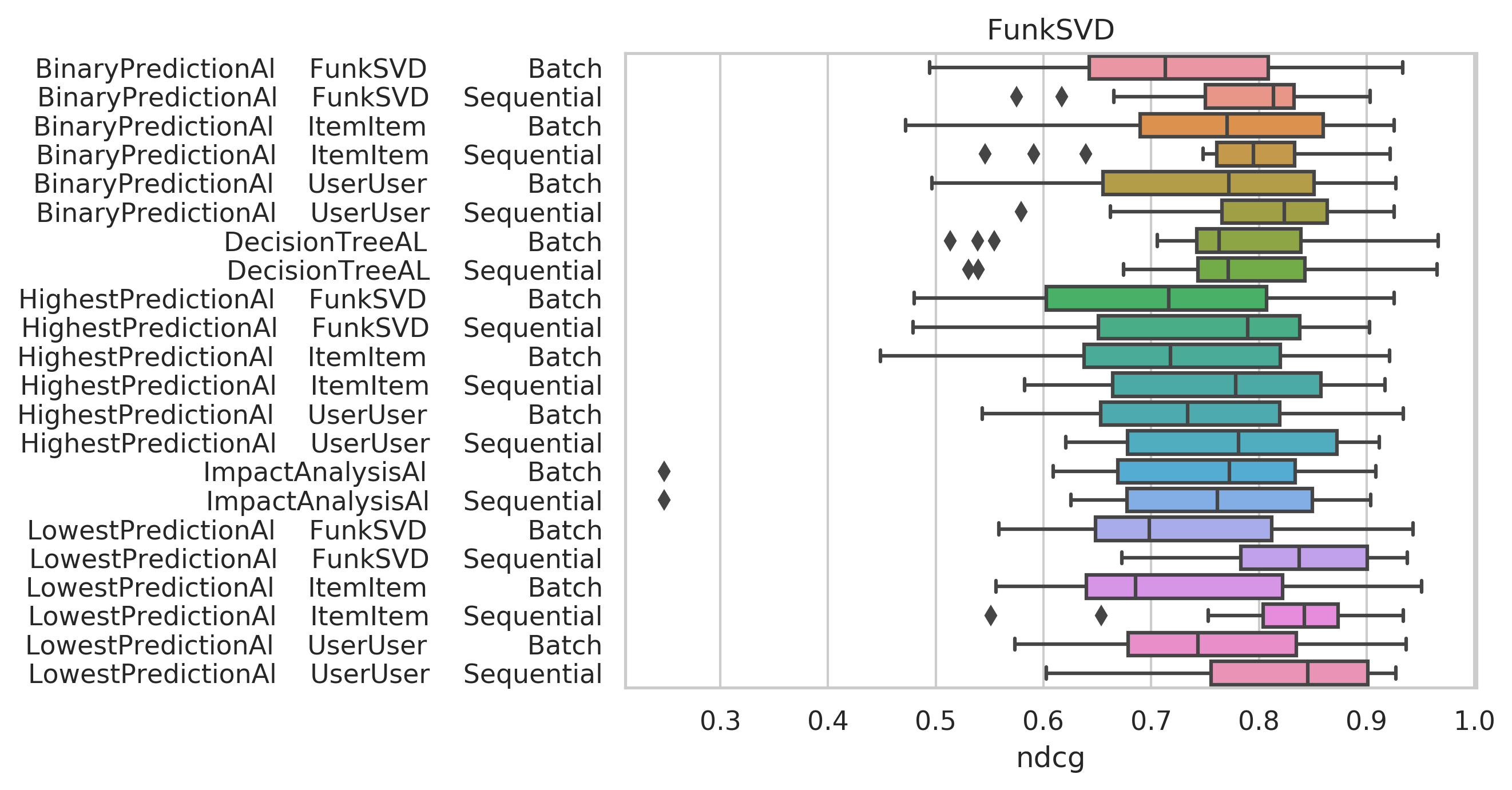}
\caption{nDCG value of the FunkSVD algorithm combined with active learners with different algorithms on the Jester dataset. Higher is better.}\label{Jester-pal-FunkSVD-ndcg}
\Description[Box-plots of the nDCG obtained for different active learners with the Jester dataset]{Box-plots of the nDCG obtained for different active learners  with the Jester dataset, showing that the median value in sequential mode is higher or the same than the median value in batch mode.}
\end{figure*}

\begin{figure*}[htbp]
\centering%
\includegraphics[width=0.98\linewidth]{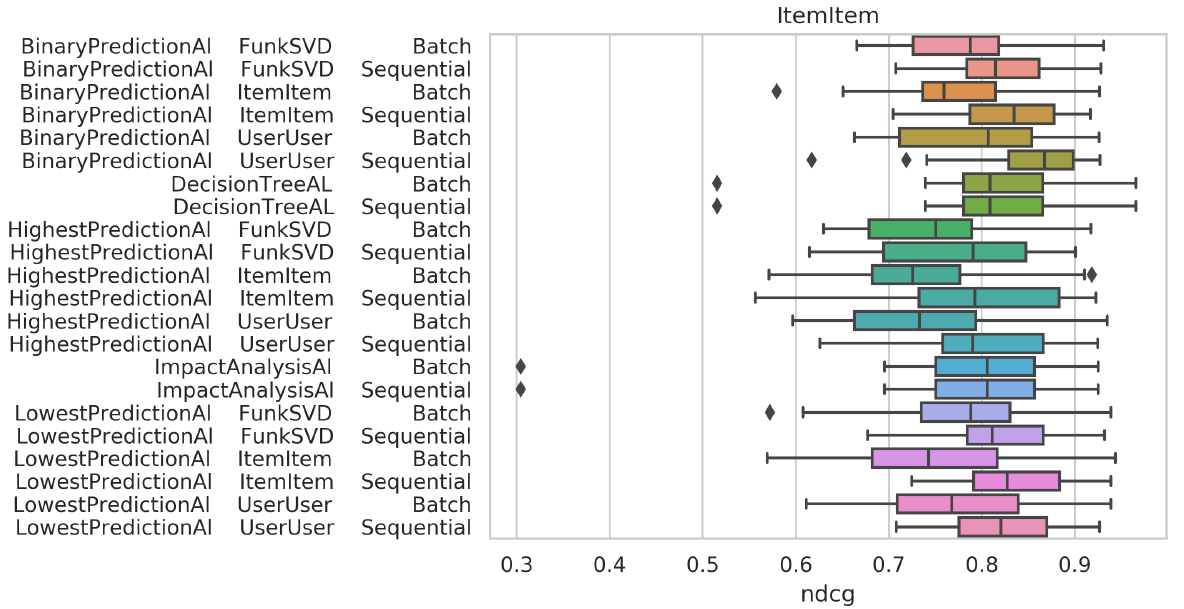}
\caption{nDCG value of item-item collaborative filtering combined with active learners with different algorithms on the Jester dataset. Higher is better.}\label{Jester-pal-ItemItem-ndcg}
\Description[Box-plots of the nDCG obtained for different active learners with the Jester dataset]{Box-plots of the nDCG obtained for different active learners with the Jester dataset, showing that the median value in sequential mode is higher or the same than the median value in batch mode.}
\end{figure*}

Figures~\ref{Jester-pal-FunkSVD-rmse},~\ref{Jester-pal-FunkSVD-precision}, and~\ref{Jester-pal-FunkSVD-ndcg} repeat this analysis for FunkSVD instead of User-User collaborative filtering as main recommendation algorithm. Figure~\ref{Jester-pal-ItemItem-ndcg} evaluates the nDCG for Item-Item Collaborative filtering on the Jester dataset. These graphs confirm the results for other recommendation algorithms. Active learning in sequential mode outperforms batch mode in terms of recommender accuracy. But the disadvantage of the sequential mode is that after each item selection, the active learner needs to re-run, which can be computational expensive. We also conclude that the differences between the active learning algorithms in terms of accuracy are small. %Also for Item-Item Collaborative Filtering as main recommendation algorithm, similar results were obtained (not shown in this paper). 

\begin{figure*}[htbp]
\centering%
\includegraphics[width=0.98\linewidth]{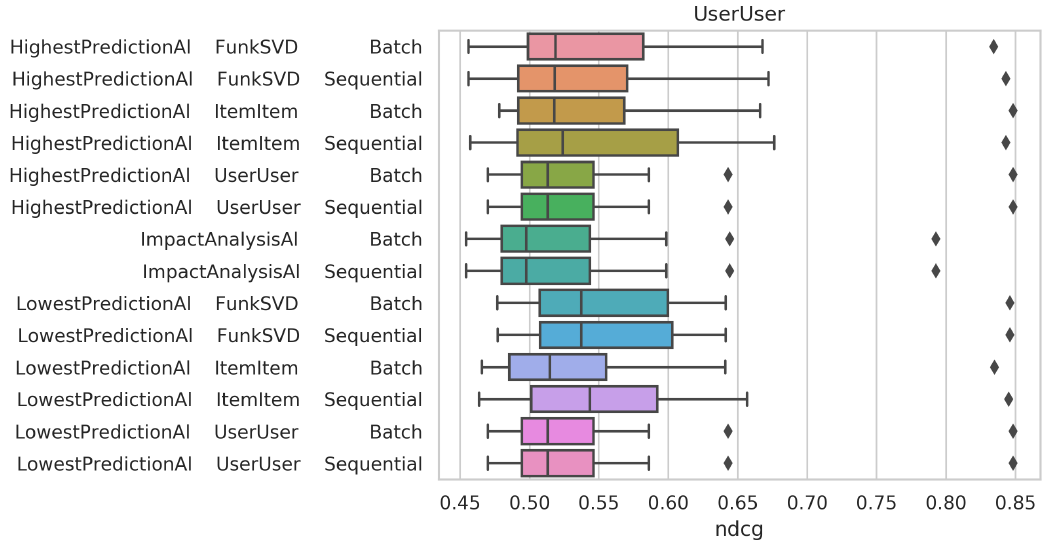}
\caption{nDCG value of user-user collaborative filtering combined with active learners with different algorithms on the Bookcrossing dataset. Higher is better.}\label{Bookscrossing-pal-UserUser-ndcg}
\Description[Box-plots of the nDCG obtained for different active learners with the Bookcrossing dataset]{Box-plots of the nDCG obtained for different active learners with the Bookcrossing dataset, showing that the median value in sequential mode is not significantly different than the median value in batch mode.}
\end{figure*}

\begin{figure*}[htbp]
\centering%
\includegraphics[width=0.98\linewidth]{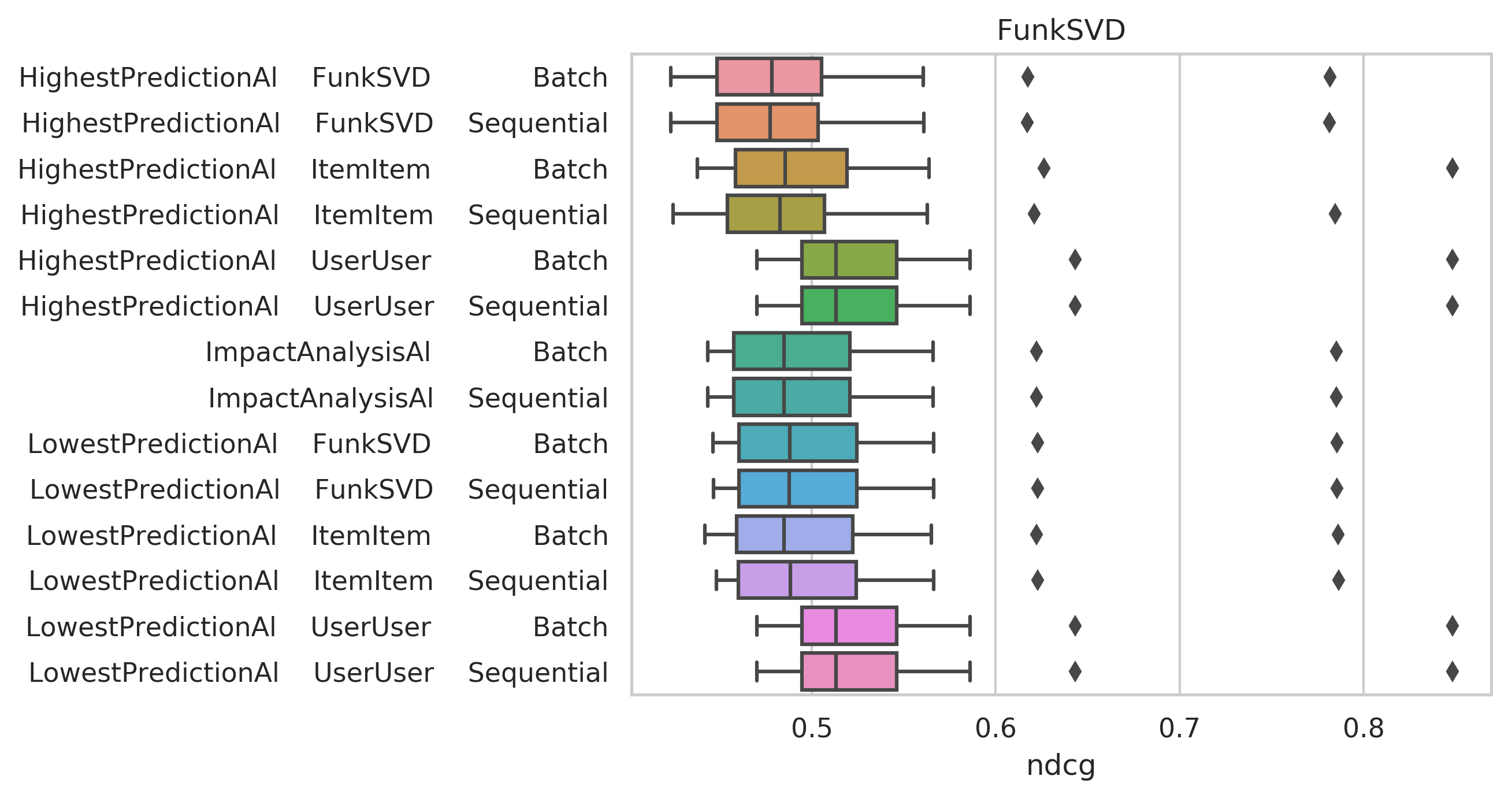}
\caption{nDCG value of FunkSVD combined with active learners with different algorithms on the Bookcrossing dataset. Higher is better.}\label{Bookscrossing-pal-FunkSVD-ndcg}
\Description[Box-plots of the nDCG obtained for different active learners with the Bookcrossing dataset]{Box-plots of the nDCG obtained for different active learners with the Bookcrossing dataset, showing that the median value in sequential mode is not significantly different than the median value in batch mode.}
\end{figure*}

\begin{figure*}[htbp]
\centering%
\includegraphics[width=0.98\linewidth]{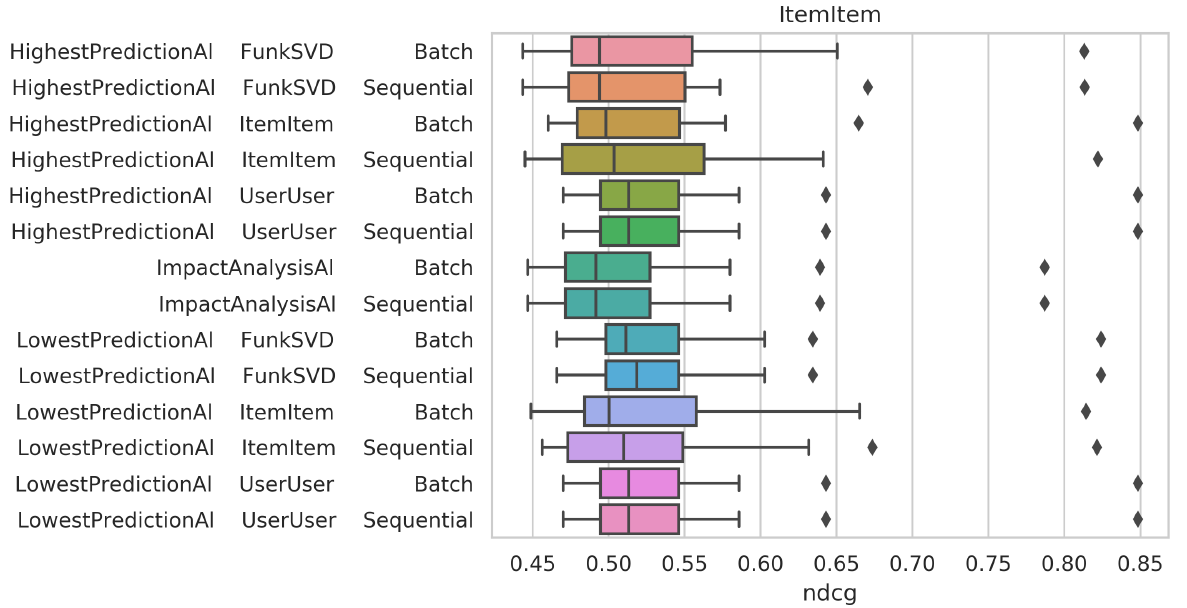}
\caption{nDCG value of item-item collaborative filtering combined with active learners with different algorithms on the Bookcrossing dataset. Higher is better.}\label{Bookscrossing-pal-ItemItem-ndcg}
\Description[Box-plots of the nDCG obtained for different active learners with the Bookcrossing dataset]{Box-plots of the nDCG obtained for different active learners with the Bookcrossing dataset, showing that the median value in sequential mode is not significantly different than the median value in batch mode.}
\end{figure*}

The analysis was also performed for the BookCrossing data set. Figures~\ref{Bookscrossing-pal-UserUser-ndcg},~\ref{Bookscrossing-pal-FunkSVD-ndcg}, and~\ref{Bookscrossing-pal-ItemItem-ndcg} show the nDCG for respectively User-User collaborative filtering, FunkSVD, and Item-Item collaborative filtering as main algorithm combined with various active learners. An interesting conclusion can be drawn here: the results obtained with active learning in sequential mode are not significantly different than with the active learning in batch mode. This was also concluded in terms of the precision and RMSE for the BookCrossing data set (not shown in this paper). The sparsity of the BookCrossing data set can be the reason that sequential mode is not better than batch mode.

%User-User algorithm%%%%%%%%%%%%%%%%%%%%%%%%%%%%%%%

%FunkSVD algorithm%%%%%%%%%%%%%%%%%%%%%%%%%%%%%%%

%Item Item algorithm

%Bookcrossing dataset%%%%%%%%%%%%%%%%%%%%%%%%%%%%%%%

%\begin{figure*}[htbp]
%\centering%
%\includegraphics[width=0.98\linewidth]{Bookscrossing-pal-FunkSVD-rmse}
%\caption{RMSE value of FunkSVD combined with active learners with different algorithms on the Bookcrossing dataset. Lower is better.}\label{Bookscrossing-pal-FunkSVD-rmse}
%\Description[Box-plots of the RMSE obtained for different active learners with the Bookcrossing dataset]{Box-plots of the RMSE obtained for different active learners with the Bookcrossing dataset, showing that the median value in sequential mode is slightly lower than the median value in batch mode.}
%\end{figure*}

\section{Conclusion and Future Work}
\label{conclusion}
Active learning algorithms carefully select informative items and ask users to rate these in order to get a more detailed view on the users' preferences and being able to generate accurate recommendations. Two modes have been compared for active learning: the batch mode, that selects all items at once, and the sequential mode, that selects items one by one and calculates in between these selections which item would be the most informative one. Five different active learning algorithms have been evaluated, and three of them were combined with three different prediction algorithms. For a dense data set, the results showed that an active learner in sequential mode enables a recommender to learn more efficiently than an active learner in batch mode. So, if users are asked to rated some items, it is better to select these items one by one, than selecting all items at once. Based on the user's ratings for the first items, the sequential active learner can better decide which subsequent items are the most informative. For a sparse dataset, no significant differences between batch and sequentially mode were obtained. For active learners that use an underlying prediction algorithm, FunkSVD turned out to be a good choice, but the accuracy differences with other predictors are small. In future work, active learners will be tested in combination with other recommendation algorithms and additional active learning algorithms will be explored.

\begin{acks}
This research received funding from the Flemish Government (AI Research Program).
\end{acks}

\bibliographystyle{ACM-Reference-Format}
\bibliography{ActiveLearningForRec}

\end{document}